\documentclass[structabstract]{aa}  
\usepackage{graphicx,amssymb,amsmath}
\usepackage{natbib}
\usepackage{txfonts}
\def\arcdeg{$^{\circ}$}
\def\H0{{\rm ~km~s^{-1}~Mpc^{-1}}}

\def\850{850$\mu$m}

\def\hii{\ion{H}{II}}

\def\.25{0.25 keV\thinspace}

\def\z2{z$\,\sim\,2$}

\begin{document}
   \title{Ultra-High Energy Cosmic Rays Detected by the Pierre Auger Observatory:}
   \subtitle{First Direct Evidence, and its Implications, that a Subset Originate in Nearby Radiogalaxies}

   \author{Neil M. Nagar
          \inst{1}
          \and
          Javier Matulich
          \inst{1}
          }

   \offprints{Neil M. Nagar}

   \institute{Astronomy Department, Universidad de Concepci\'on, Concepci\'on, Chile \\ 
              \email{nagar@astro-udec.cl,jmatulich@udec.cl}
             }

   \date{Received May 5, 2008; accepted June 16, 2008}

  \abstract
   {The Pierre Auger Collaboration has reported 27 Ultra-High
    Energy Cosmic Ray Events (UHECRs) with energies above 
    56$\times 10^{18}$ eV (56~EeV) and
    well determined arrival directions as of 2007 August 31.  
    They find that the arrival directions of these UHECRs are not 
    isotropic, but instead appear correlated with the positions of 
    nearby Active Galactic Nuclei (AGNs)
    from the catalog of V{\'e}ron-Cetty \& V{\'e}ron.
   }
   {Our aim was to determine the sources of these UHECRs by comparing 
    their arrival directions with more complete and/or comprehensive
    astronomical source catalogs.
   }
   {We have cross correlated the arrival directions of the 
    UHECRs with the positions of
    supernovae, radio supernovae, galaxies, 
    active galaxies, radiogalaxies, and clusters of galaxies,
    all at distances within $\sim$200~Mpc.
   }
   {Four (eight) of the 27 UHECRs with energy greater than 56~EeV 
    detected by the Pierre Auger Observatory have arrival directions 
    within 1.5\arcdeg\ (3.5\arcdeg) of the extended ($\ge$ 180~kpc)
    radio structures of nearby radiogalaxies or the single nearby BL~Lac
    with extended radio structure.
    Conversely the radio structures of three (six) of all ten nearest extended 
    radiogalaxies are 
    within 1.5\arcdeg\ (3.5\arcdeg) of a UHECR; three of the remaining four radiogalaxies
    are in directions with lower exposure times. This correlation between
    nearby extended radiogalaxies and a subset of UHECRs is significant at the 
    99.9\% level. 
    For the remaining $\sim$20 UHECRs, an isotropic distribution cannot be
    ruled out at high significance. 
    The correlation found by the Auger Collaboration
    between the 27 UHECRs and AGNs in the 
    V{\'e}ron-Cetty \& V{\'e}ron catalog at D$\lesssim$71~Mpc 
    has a much lower significance when one considers only the $\sim$20 UHECRs 
    not `matched' to nearby extended radiogalaxies.
    No clear correlation is seen between UHECRs and 
    supernovae, supernova remnants, nearby galaxies,
    or nearby groups and clusters of galaxies. 
   }
   {Nearby extended
    radiogalaxies are the most likely source of at least some
    UHECRs detected by the Pierre Auger Observatory. The remaining
    UHECRs are not inconsistent with an isotropic distribution; their
    correlation to nearby AGNs is much less significant than earlier
    estimated. 
    This is the first direct observational proof that
    radio galaxies are a significant source of UHECRs.
    The primary difference between the UHECR detections at the 
    Pierre Auger Observatory and previous experiments, e.g. AGASA,
    may thus primarily be that the Southern Hemisphere
    is more privileged with respect to nearby radiogalaxies with
    highly extended radio jets and lobes.
    }

   \authorrunning{Nagar \& Matulich}
   \titlerunning{UHECRs from Radiogalaxies}
   \maketitle
%

\section{Introduction}
\label{secintro}

Ultra-High Energy Cosmic Rays (UHECRs) are protons or light
nuclei with energies greater than about 10$^{19}$ eV (10 EeV). 
When these highly energetic particles enter the earth's atmosphere
they produce a shower of secondary particles and excite
the Nitrogen molecules in the atmosphere; both effects can be
detected from the ground. Recent advances in ground shower
detectors allow a precise measurement of both the initial energy and the
arrival direction of the UHECR above the earths atmosphere,
e.g. AGASA \citep{taket99} and 
the Pierre Auger Observatory \citep{abret04,demet07}.
While UHECRs are rare they have a unique astronomical potential:
UHECRs with energies $\gtrsim$50~EeV are not expected to be 
deflected significantly by Galactic or intergalactic magnetic 
fields. Their measured arrival directions can therefore be traced 
directly back to the originating source within the observatory 
measurement uncertainties.

Suspected sources of high energy cosmic rays include 
core collapse Supernovae (SN II), 
Supernova Remnants,
Pulsars,
Active Galaxies - especially radio jets and radio lobes, and
gamma rays bursts (for reviews and lectures see
\citet{kat08,hil99} and references therein).
In these sources, the cosmic rays are posited to undergo
successive accelerations via scattering off energetic
charged particles and/or in shocks. 
More exotic explanations include an origin in dark matter anhilation.

The main energy loss for UHECR propagating over cosmological 
distances is expected to be pion-production, the so called 
Greisen Zatsepin Kuzmin  (GZK) effect \citep{gre66,zatkuz66}.
In this process, a UHECR interacts with a CMB photon and
loses an estimated $\sim$30\% of its energy. 
Both the energy loss and the mean free path between interactions
depend on several details; energy loss length predictions vary 
between 20 and 100~Mpc (see e.g. \citet{sta04},\citet{kat08},PA08). If the
GZK effect is at work then UHECRs with energy above 
$\sim$50~EeV are expected to originate primarily in sources closer
than this predicted energy loss length. 

The Pierre Auger Observatory (hereafter PAO) in Argentina \citep{abret04,demet07}, 
an array of 1600 Cerenkov detectors spread over 3000 km$^2$ plus 
six optical telescopes, is designed
to measure arrival directions and energies of cosmic rays via 
their secondary particle showers and their atmospheric flouresence.
Latest details on the instrumentation and exciting science results 
being produced by the Auger Observatory and Collaboration can be
found at the Auger web-site http://www.auger.org.
The array has been in partial operation since January 2004.
The Auger collaboration has
reported 81 events between 2004 January 1 and 2007 August 31  
with reconstructed energies above 40~EeV and
zenith angles smaller than 60\arcdeg; of these, 27 have energies
above 56~EeV \citep[][hereafter PA07,PA08]{abret07,abret08}.
The origins of the latter 27 UHECRs are the focus of this paper.

The Auger Collaboration found that the arrival directions of
the 27 UHECRs with energies above 56~EeV are 
not isotropic at a 99\% significance level (PA07,PA08).
Instead, they found that the arrival directions are 
correlated with the positions of AGNs within
$\sim$71~Mpc (PA07,PA08,\citet{molet07}). 
For this analysis they used
all AGNs (and galaxies with \hii\ nuclei) 
in the catalog of V{\'e}ron-Cetty \& V{\'e}ron, 12th
Edition \citep{verver06}. 
In their statistical correlation of all 81 UHECRs at
energy above 40~EeV with the V{\'e}ron-Cetty \& V{\'e}ron catalog of AGNs, 
PA07,PA08 found that the highest correlation between UHECRs and 
AGNs was obtained for a maximum AGN distance of 71~Mpc, 
an angular separation of 3.2\arcdeg\ between ``matched'' AGNs and UHECRs, 
and an energy threshold for UHECRs of 56~EeV.
This maximum distance, 71~Mpc, is in line with the expectations of
the GZK effect. 
\citet{goret07}, in an astro-ph comment, disagreed with this
finding, pointing out that weighting each AGN with its distance
would predict a very different distribution of UHECRs than that found
by the PAO.

Our interest in this topic was
provoked when a quick examination of the 27 UHECR events
showed that the direction of at least five of the 27 
UHECR events were directly in the line of sight to 
nearby extended radiogalaxies.
In this article, we attempt a more comprehensive correlation
between the arrival directions of the 27 UHECRs with energies
above 56~EeV and catalogs of potential sources of UHECRs,
both Galactic and Extragalactic.
Sec. \ref{secdata} describes the data used in our study, 
Sec. \ref{secres} describes the principle results obtained, 
and Sec. \ref{secdis} contains a 
brief discussion and conclusions of our study.
Distances to galaxies are calculated using 72 $\H0$, except for 
relatively nearby galaxies for which we use
distances as referenced.

\section{Data}
\label{secdata}

Positions and energies of the 27 UHECRs detected by the PAO
with energies $\geq$~56~EeV 
were taken from PA08. The figures in this publication are slightly
different from those of the Auger collaboration: PA07 and PA08 
used the Aitoff projection (an equidistant projection with 2:1 aspect ratio), 
while we use the Aitoff-Hammer projection (an equal area projection). The
latter is more commonly used in astronomy and is typically referred to as
simply the `Aitoff' projection, e.g. the IDL procedure aitoff.pro.
Positions and energies of the UHECRs detected by AGASA
are taken from \citet{hayet00} which is an updation of the data listed
in \citet{taket99}. 

Our catalogs of astronomical sources are drawn from various
publications and web-based source lists. 
In all cases we attempted to assemble as complete a
sample as possible. Further, following the results of PA07, PA08,
and the expectations from the GZK effect, we considered
only objects within a distance of $\sim$200~Mpc. For matches
to the PAO UHECRs we consider only those
sources which rise above elevation 30\arcdeg\ at PAO
since the PAO UHECR list includes only UHECRs detected above this
elevation.

Our list of extragalactic
supernova remnants, galaxies, and galaxy clusters
were drawn from NED \citep{madet92} using the `All Sky Search' option.
The list of Gamma Ray Bursts (GRBs) was taken from the Gamma-ray Burst Real-Time
Sky Map{\footnote{http://grb.sonoma.edu}}.
For Galactic Supernova Remnants we used the list maintained by D. 
Green{\footnote{http://www.mrao.cam.ac.uk/surveys/snrs/snrs.data.html}}.
Extragalactic Radio Supernova were taken from Kurt
Weiler's list at the web-site of the Naval Research
Laboratory{\footnote{http://rsd-www.nrl.navy.mil/7213/weiler/kwdata/RSNtable.txt}}.
For extragalactic supernova, we used the Sternberg Astronomical Institute
(SAI) Supernova 
Catalog{\footnote{http://www.sai.msu.su/sn/sncat/}} maintained by
Tsvetkov et al.
In the case of radiogalaxies and radio sources we used
several surveys and catalogs, including NVSS \citep{conet98},
SUMMS \citep{bocet99}, and NED. The list of extragalactic jets is based
on that compiled by \citet{liuzha02}; to this list we added data on the galaxies
as given in NED, and re-measured the total flux and total extent 
of the radio emission from NVSS and SUMMS maps. 
The total extent of the radio emission therefore includes both
radio jets and any radio lobes. This extent is typically referred to as
the Largest Angular Size (LAS) or Largest Linear Size (LLS). In this work
we added the LLS of the individual jets and lobes as scalars instead of vectors
in order to discount the effects of jet and lobe bending.

To recreate the results of the Auger Collaboration (PA07,PA08), we
used the 12th edition of the \citet{verver06} catalog of AGNs.
The list of AGNs (agn.dat) in this catalog includes nuclei with
\hii\ spectra which were previously misclassified as AGNs. These
should ideally be left out of the correlation analysis. From the
AGN statistics quoted in PA07,PA08, it appears 
that the Auger Collaboration included these \hii\ nuclei in their 
analysis, so for consistency we do the same. 
Additionally, the list of QSOs (qso.dat)
lists seven galaxies with redshift=0. Four of these are 
truly nearby galaxies, and the other three are probably high redshift 
quasars. Again, for consistency with PA07 and PA08 we keep the
latter three in our analysis. 

We have also used the Fanaroff-Riley (FR) classification for radiogalaxies
\citep{fanril74}. This division is based on whether the jets increase
(FR~II) or decrease (FR~I) in brightness with increasing distance
from the nucleus. A clear division in radio luminosity is found between the
two classes, with FR~IIs being more luminous in the radio 
than FR~Is \citep{oweled94}.

\section{Results}
\label{secres}

A comparison of the arrival directions of the 27 UHECRs
and the positions of nearby galaxies with radio jets 
\citep[all galaxies listed in][]{liuzha02}
is shown in Fig.~\ref{figjetned}. 
We have divided the galaxies with radio jets into three redshift
bins: ``nearby'' (D $\leq$ 75~Mpc), ``intermediate'' (75~Mpc $<$ D $\leq$ 150~Mpc),
and ``distant'' (150~Mpc $<$ D $\leq$ 210~Mpc). In the closest
redshift bin, D $\leq$ 75~Mpc, 
we distinguish between galaxies with radio structures more extended than 180~kpc
(``extended'') and those with radio structures less extended than 180~kpc 
(``compact'').
We calculated the total extent of the radio structure of any given galaxy by 
summing the lengths of the two radio jets and/or lobes as scalars instead of vectors. 
The total radio extent includes radio jets and any radio lobes.

There is no intrinsic reason to use a radius $\psi=$~3.5\arcdeg\ for
a ``match'' between UHECR arrival direction and the position of an
astronomical source. The instrumental resolution of the PAO
is about 1\arcdeg\ at these UHECR energies, to which one
must add a small angle -- perhaps up to a few degrees \citep{sta04} --  to account
for the deflection of the UHECR by magnetic fields. 
The best correlation between UHECRs and nearby AGNs found by PA07, PA08  
is based on considering $\psi \sim$3.2\arcdeg as a match. However, as we
show later (e.g. Fig.~\ref{figscanb}), the value of $\psi$ in the 
analysis of PA07,PA08 should really be $\psi \sim$ 3.1\arcdeg--4.3\arcdeg.
For the above reasons, we consider two values of $\psi$ for a match: 
$\psi=$~1.5\arcdeg\ and $\psi=$~3.5\arcdeg. 
The distance limit for `nearby' galaxies - 75~Mpc - is similar to
the value d$_{\rm max} \sim$ 71~Mpc found by PA07 and PA08.
The division between `compact' and `extended' radio structures 
- 180~kpc - is the one `new' free parameter in our analysis. We note that
this value is close to the minimum linear size of the radio structures of a 
typical radio galaxy \citep[e.j.][]{haret98}.

Of all 27 UHECRs with energy above 56~EeV detected by the PAO
four (eight) are within $\psi=$~1.5\arcdeg\ ($\psi=$~3.5\arcdeg)
of the radio structures of nearby extended radiogalaxies in the 
field of view of the PAO.
Conversely, of all ten nearby extended radiogalaxies in the
field ov view of the PAO three (six) 
have radio structures which
can be matched to within 1.5\arcdeg\ (3.5\arcdeg) of a UHECR event. Of the remaining 
four `unmatched' radiogalaxies, three - NGC~1316, NGC~4261 and NGC~4760 - fall outside
the area of maximum exposure at the PAO.
For the intermediate and distant redshift bins
there is little match between extragalactic radio jets and
UHECRs; the only two matches within 3.5\arcdeg\ are CGCG~403-019, a 
Seyfert/BL~Lac
at 112~Mpc, and Mrk~612, a Seyfert galaxy at 85~Mpc.

The match between a UHECR and CGCG~403-019 deserveres
furter comment.
CGCG~403-019 is a BL~Lac with a Seyfert~1 spectrum 
\citep{verver93}. In the 12th AGN Catalog of
\citet{verver06} it is listed by its alias
PKS~2201+04 and is classified as a Seyfert 1.
It is a poweful radio source with a core-jet 
structure with total extent $\sim$6{\arcmin} or
$\sim$190~kpc \citep{ulvjoh84,lauet93}. Within 
the unified scheme of radiogalaxies, BL~Lacs are 
posited to be pole-on FR~Is, implying a much
larger deprojected radio extent. 
We therefore consider CGCG~403-019 as an `intermediate'
distance `extended' radiogalaxy.
Further, CGCG~403-019 is unique in being the
nearest BL~Lac with extended radio structure.
The 12th AGN Catalog of \citet{verver06} includes only 
three confirmed BL~Lacs - V~Zw~331, RXS J05055+0416, and
TEX 0554+534 - and only one probable BL~Lac - 
RXS J21231-1036 - at D $\leq$ 150~Mpc. All four show
compact radio emission in NVSS radio maps.

The Auger Collaboration noted that the region around Cen~A has 
an unusually large number of UHECRs; this is also the area
with the maximum number of nearby extended radiogalaxies.
As seen in Fig.~\ref{figsjetned}, the two UHECRs centered on
Cen~A can be explained as originating in either of Cen~A,
NGC~5090 (D=48~Mpc; total radio extent $\sim$187~kpc), or even
PKS~1308-441 (D=211~Mpc; total radio extent $\sim$1140~kpc).
The event just further south can be matched to the
southern jet of Cen~A, or to the relatively distant PKS~1308-441.
Cen~B (D=54~Mpc; total radio extent $\sim$250~kpc) is almost
perfectly centered on a UHECR, and
WKK~4552 (D=67~Mpc; total radio extent $\sim$360~kpc)
is within $\sim$3\arcdeg\ of another UHECR. The
UHECR just to the north of Cen~A is bracketed by
IC~4296 (D=52~Mpc; total radio extent $\sim$1560~kpc)
and
ESO~443-G024 (D=71~Mpc; total radio extent $\sim$60~kpc). The
very extended radio structure of the former actually intersects with
the 3.5\arcdeg\ circle around the UHECR event. 

\begin{figure}
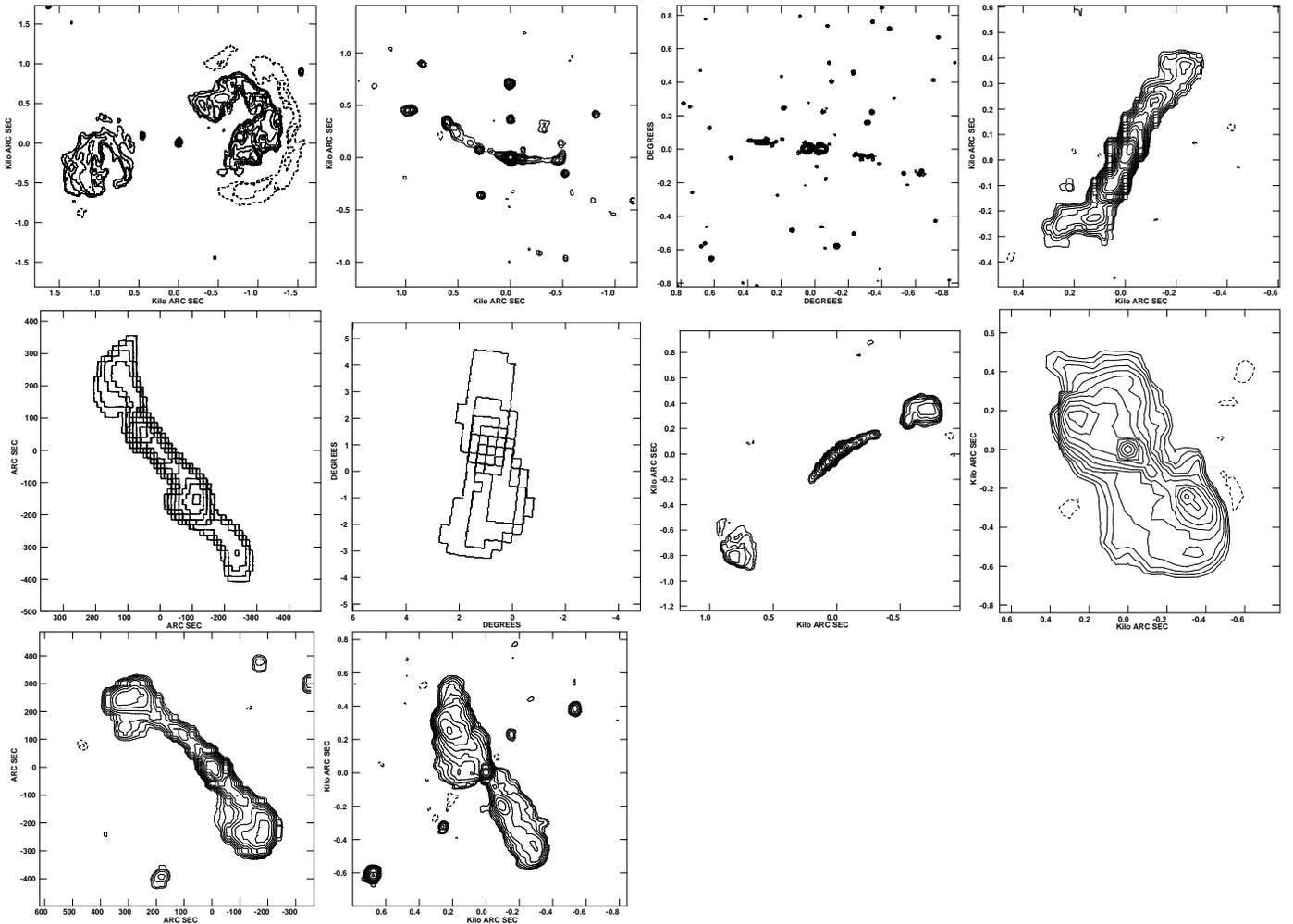

\resizebox{\textwidth}{!}{
\includegraphics[bb=46 172 565 670,clip]{N1316SUMMS.PS}
\includegraphics[bb=36 172 565 670,clip]{N3557NVSS.PS}
\includegraphics[bb=36 172 565 670,clip]{N4261NVSS.PS}
\includegraphics[bb=36 172 565 670,clip]{N4760NVSS.PS}
}

\resizebox{\textwidth}{!}{
\includegraphics[bb=36 157 565 690,clip]{N5090SUMMS.PS}
\includegraphics[bb=36 166 565 675,clip]{CENA408.PS}
\includegraphics[bb=36 173 565 672,clip]{IC4296NVSS.PS}
\includegraphics[bb=36 157 565 690,clip]{CENB.PS}
}
\resizebox{3.55in}{!}{
\includegraphics[bb=36 174 565 670,clip]{N7626NVSS.PS}
\includegraphics[bb=36 174 565 670,clip]{W4552SUMMS.PS}
}
\caption{The radio morphologies of the radiogalaxies in Table~1.
Unless otherwise specified, the radio maps are from NVSS.
From left to right, the contour maps are 
top row: NGC~1316 (SUMMS), NGC~3557, NGC~4261, NGC~4760;
middle row: NGC~5090 (SUMMS), Cen~A (408~MHz), IC~4296, Cen~B (SUMMS);
bottom row: WKK~4452 (SUMMS), NGC~7626. 
Radiogalaxies which match
the positions of UHECRs (middle and bottom rows) are preferentially
those with intermediate FRI-FRII type characteristics. Those not
matched to positions of UHECRs (top row) are FR~Is or in the case
of NGC~1316, a `relaxed double'.
}
\label{figcontour}
\end{figure}

Are the nearby extended radiogalaxies matched to PAO UHECRs different 
from those not matched to PAO UHECRs? 
A comparison of the relevant properties of the ten nearest
extended radiogalaxies is given in Table~1, and their 
radio contour maps in Fig.~\ref{figcontour}. Table~1 also lists
the relevant radio properties of CGCG~403-019 and Mrk~0612, the
two intermediate distance AGNs which match to a UHECR. 
There is some difference in the combination of two factors: 
the radiogalaxies in the direction to UHECRs include all of
the subset which have radio jet morphologies closer to FR~II, 
and are all in the area where
the PAO has obtained a higher exposure
time. In any case, given the small number statistics, i.e. typically
1$\pm$1 UHECR events per matched radiogalaxy, it is not
surprising that UHECRs are not yet detected towards the
remaining four radiogalaxies. Clearly more UHECRs are needed
to form any firm conclusion.

Fig.~\ref{figcluster} compares the arrival directions of 
UHECRs with the distribution of nearby groups and clusters of galaxies.
While there is an over-density of galaxies, clusters of
galaxies and UHECRs around the region of Cen~A, the overall
distributions of nearby galaxies and nearby galaxy clusters
do not correlate with the arrival directions of UHECRs.
This argues against an origin of UHECRs in material associated
with large mass haloes, e.g. Dark Matter.
The more detailed analysis of \citet{goret07} also predicts
a different distribution of UHECRs when the distances to 
galaxies and clusters are taken into account.

We have also compared the arrival directions of the 
27 UHECRs with the positions of Galactic and
nearby extragalactic supernovae and supernova
remnants,
of nearby extragalactic radio supernovae, of
nearby BL~Lacs, and of gamma-ray bursts.
In all cases there appears to be no
more than a random overlap.

\begin{figure}
\includegraphics[width=3.2in,clip]{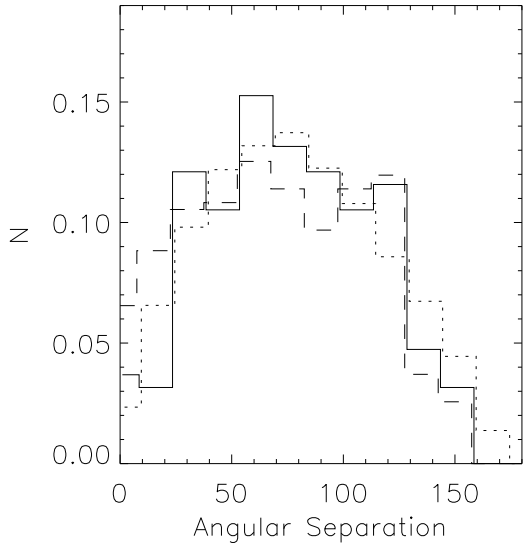}
\caption{Histograms of the angular separation between UHECRs detected
by the PAO. 
The dashed line is for all 27 UHECRs with energy above 56~EeV. 
The solid line is for the subset of 19 UHECRs which are not
matched to nearby radiogalaxies with extended radio structures (see
text). The dotted line is the expectation for a uniform
distribution of the 27 UHECRs given the exposure time
weighting in declination for the PAO. 
The histograms are offset by a few degrees
in $x$ for clarity.
}
\label{fighist}
\end{figure}

The analysis of PA07 and PA08 considered a single source
population for the UHECRs. How would their results
change if one were to consider multiple source populations,
at least one of which is nearby extended radiogalaxies?
Fig.~\ref{fighist} compares the histograms of the
angular distribution of the 27 UHECRs with energies above
56~EeV, the subset of 19 UHECRs not towards nearby 
extended radiogalaxies,
and the expectations from a random distribution of 27 UHECRs.
The deviation from isotropy discussed by 
PA07 and PA08 (dashed line in the figure)
comes primarily from the clustering of events
around Cen~A. Deleting the UHECRs which lie within 3.5\arcdeg\ of
nearby radiogalaxies (solid line in the figure) gives a distribution
in which isotropy cannot be ruled out at high significance.

\begin{figure}
\resizebox{5in}{!}{
\includegraphics[width=2.5in,clip]{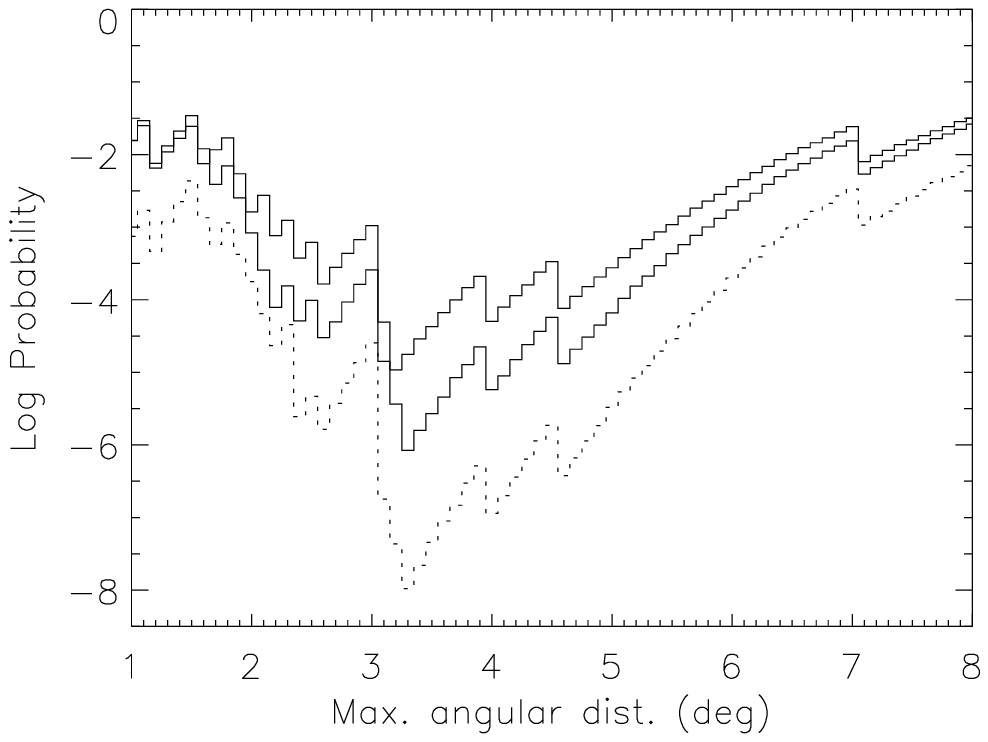}
\includegraphics[width=2.5in,clip]{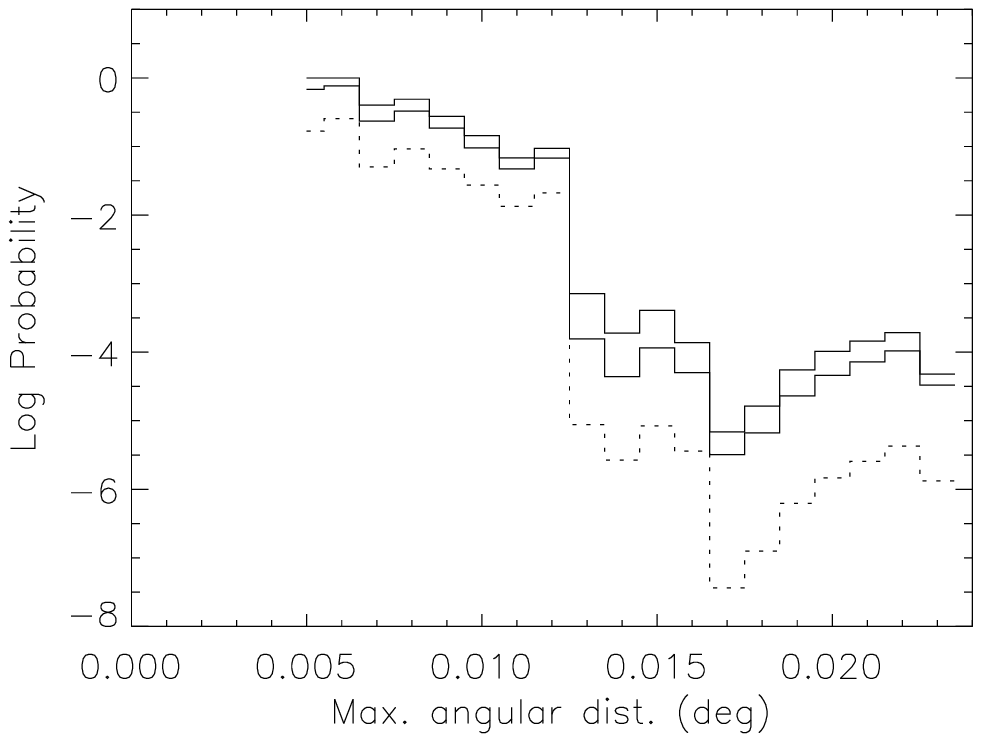}
}
\caption{Probability of the null hypothesis (isotropic distribution)
vs. maximum angular distribution $\psi$ (left panel) and maximum
AGN redshift z$_{\rm max}$ (right panel). 
In this figure we followed the analysis of PA07,PA08, i.e.
correlating UHECRs to \textit{all} galaxies at z $\le$ 0.024 in
the AGN Catalog 12th ed. \citep{verver06}.
The dashed lines are the results of using all 27 UHECRs, i.e. the same
plots as in PA07 and PA08. 
The middle solid line is the result of deleting all 4 UHECRs within 1.5\arcdeg\
of the radio structures of nearby extended radiogalaxies
 and the upper solid line is the result of
deleting all 8 UHECRs within 3.5\arcdeg\ of the radio structures of nearby extended radiogalaxies
(see text).
}
\label{figscana}
\end{figure}

\begin{figure}
\resizebox{5in}{!}{
\includegraphics[width=2.5in,clip]{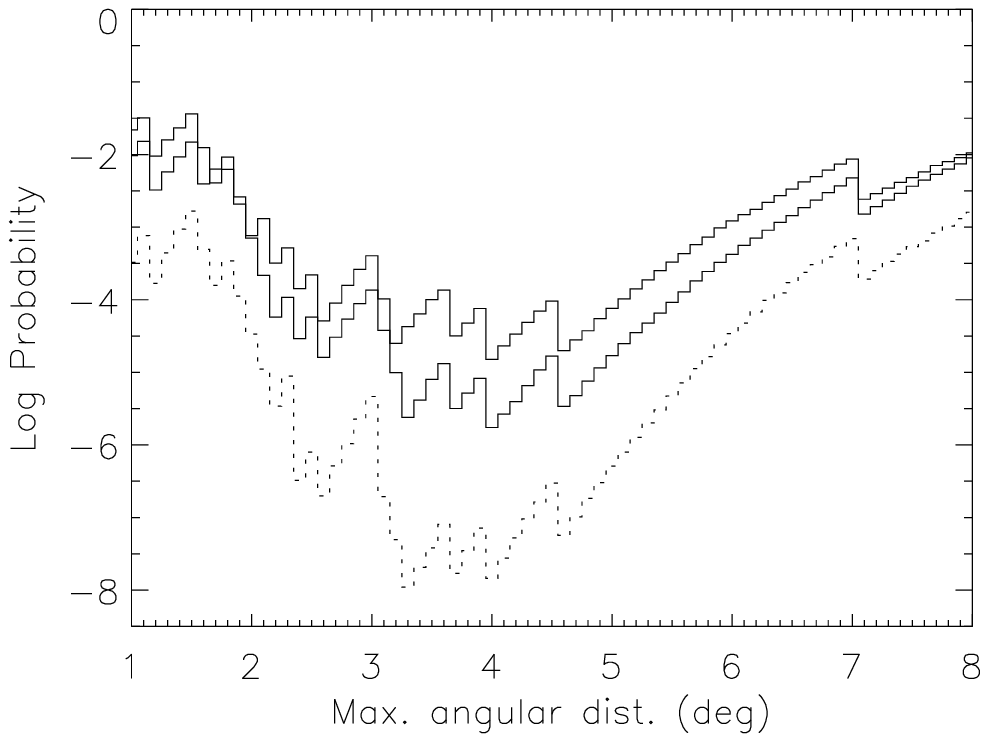}
\includegraphics[width=2.5in,clip]{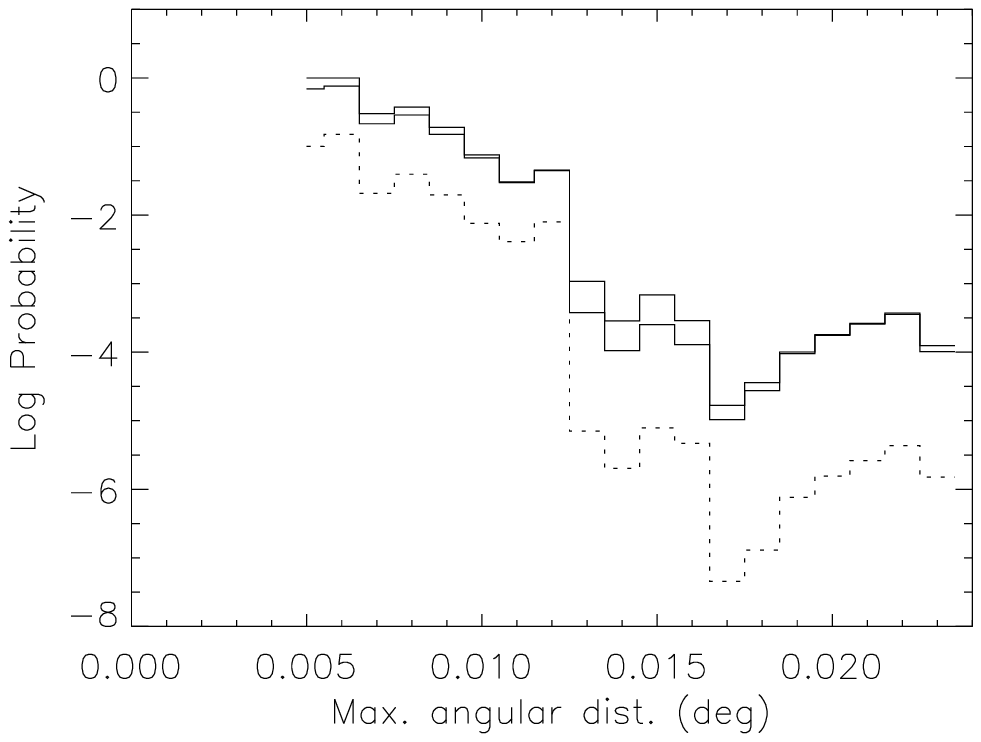}
}
\caption{The same as Fig.~\ref{figscana}, but this time, of
all galaxies listed in the AGN Catalog 12th ed., we do not
consider galaxies with H~II type nuclei and the three high redshift
QSOs which are listed as z=0 sources
(see text). The main difference from Fig.~\ref{figscana} is that the
minimum value for the null hypotesis in $\psi$ is now over a broader range: 
$\psi$ $\sim$ 3.1\arcdeg to 4.3\arcdeg.
}
\label{figscanb}
\end{figure}

The Auger Collaboration cross correlated the positions of the 27 UHECRs
with the AGNs in the 12th catalog of \citep{verver06} by 
varying the maximum AGN distance (z$_{\rm max}$), 
the angular separation between AGN and UHECR to be considered a match
($\psi$), and the threshold energy of UHECRs. 
They then looked for the minimum cumulative probability of chance coincidences.
Note that they
apparently included \hii\ nuclei from the catalog in their analysis. 
In this process (see PA08 for details) they found that
the best match was found for z$_{\rm max}$ = 0.017 (D $\sim$ 71~Mpc)
and $\psi$ = 3.2\arcdeg. 
We repeated their cumulative probability scans in
the parameters z$_{\rm max}$ and $\psi$; 
we were unable to scan in energy as we do not have
access to the position and energy data on the UHECRs  with energies less than
56~EeV. The resulting scans are shown in Fig.~\ref{figscana}. The dashed line
shows the result if all 27 UHECRs are used, i.e. the result
obtained by PA08. In this case 19 of the 27 UHECRs match to some
AGN or \hii\ nuclei in the catalog.
The middle solid line of each panel shows the result of considering only the 23 UHECRs 
remaining after deleting the four UHECRs which best match (within 1.5\arcdeg)
the radio structures of nearby extended radiogalaxies
(2 UHECRs for Cen~A, and one each for Cen~B and NGC~7626). In this case
13 of the 23 UHECRs match to some AGN or \hii\ nuclei in the catalog.
The upper solid lines show the results of considering only the 19 UHECRs remaining
after deleting the four UHECRs mentioned above, plus the UHECRs 
matched within 3.5\arcdeg\ to the South jet of Cen~A, to CGCG~403-019, to
WKK~4452, and to IC~4296.
In this case 11 of the 19 UHECRs match to some
AGN or \hii\ nuclei in the catalog.
While the overall shape of the scans in $\psi$ and z$_{\rm max}$ 
remain roughly the same, the significance of the results decrease
significantly. In the case of the scan in $\psi$, the minimum at
around 3.2\arcdeg\ is no longer a clear and unique minimum.
We repeated the above exercise after making two modifications
to the AGN list: we deleted all galaxies with H~II type nuclei 
(50 galaxies at z $\le$ 0.024)
and the three probable high redshift QSOs which are listed as having
z=0 in the AGN Catalog 12th. ed.. The results are shown in Fig.~\ref{figscanb}. 
The main difference with the previous figure is the broader valley
in the $\psi$ scan. Instead of a clearly defined minimum at $\psi$ = 3.2\arcdeg 
one now has a greater range $\psi \sim$ 3.2\arcdeg--~4.3\arcdeg.

\section{Discussion \& Concluding Remarks}
\label{secdis}

The arrival directions a subset of 
the 27 UHECRs with energies above 56~EeV detected by the
PAO are statistically most closely related
to nearby radiogalaxies with extended radio jets and/or lobes.
It is thus likely that nearby (due to the GZK effect) 
radiogalaxies with extended (radio extent $\ge$ 180~kpc), two sided 
jets and lobes
are the source population of a subset of the UHECRs detected by the
PAO. Additionally, there is weak evidence
that hosting a jet with morphology closer to FR~II makes a 
radiogalaxy more likely
to be a UHECR source. Interestingly, all these factors are
consistent with the lack of UHECR detections towards M~87.

\begin{figure}
\includegraphics[width=4.5in,clip]{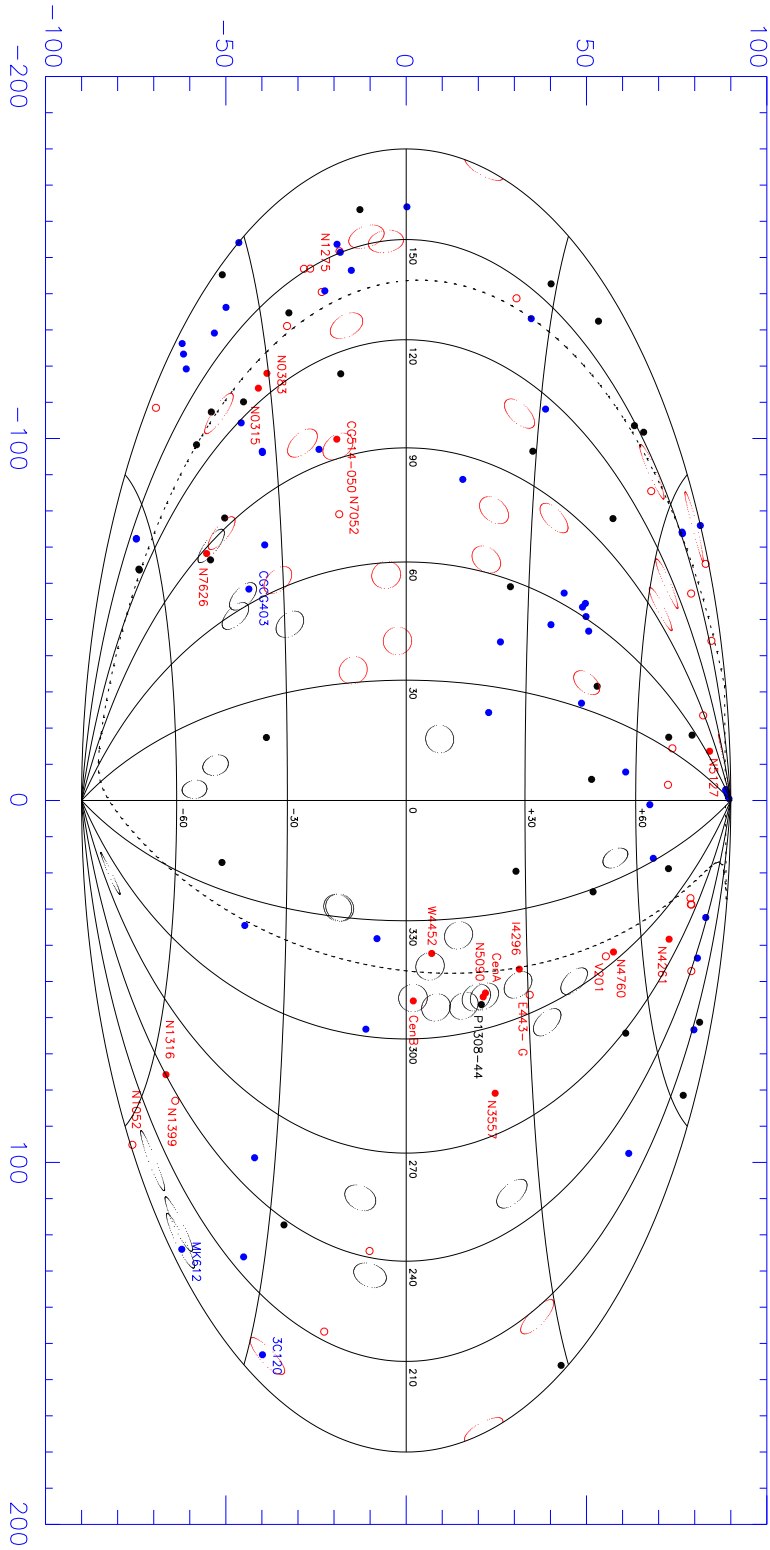}
\caption{A comparison of the arrival directions of the 27 UHECRs with
the positions of galaxies hosting radio jets.
PAO UHECR events with energies above 56~EeV are shown as 
black circles with radius 3.5\arcdeg. AGASA UHECR events with energies
above 56~EeV are shown as red circles 
with radius 3.5\arcdeg.
The colour symbols mark the positions of galaxies with radio jets at
D $\leq$ 75~Mpc (red),
75~Mpc $<$ D $\leq$ 150~Mpc (blue), and
150~Mpc $<$ D $\leq$ 210~Mpc (black).
For galaxies at D $\leq$ 75~Mpc, filled symbols are used for galaxies with 
radio structures more extended
than 180~kpc, and open symbols for galaxies with radio structures less
extended than 180~kpc. Galaxy names are marked in some of the
less crowded regions in the plot, in the same color as the corresponding
symbol. The SuperGalactic plane is marked by the dashed line.
In this and following figures, we use an Aitoff Hammer Projection.
The orientation and labelling of the longitude axis follows PA07 and PA08 
for easier comparison.
}
\label{figjetned}
\end{figure}

\begin{figure}
\includegraphics[bb=260 230 330 300,width=3in,angle=90,clip]{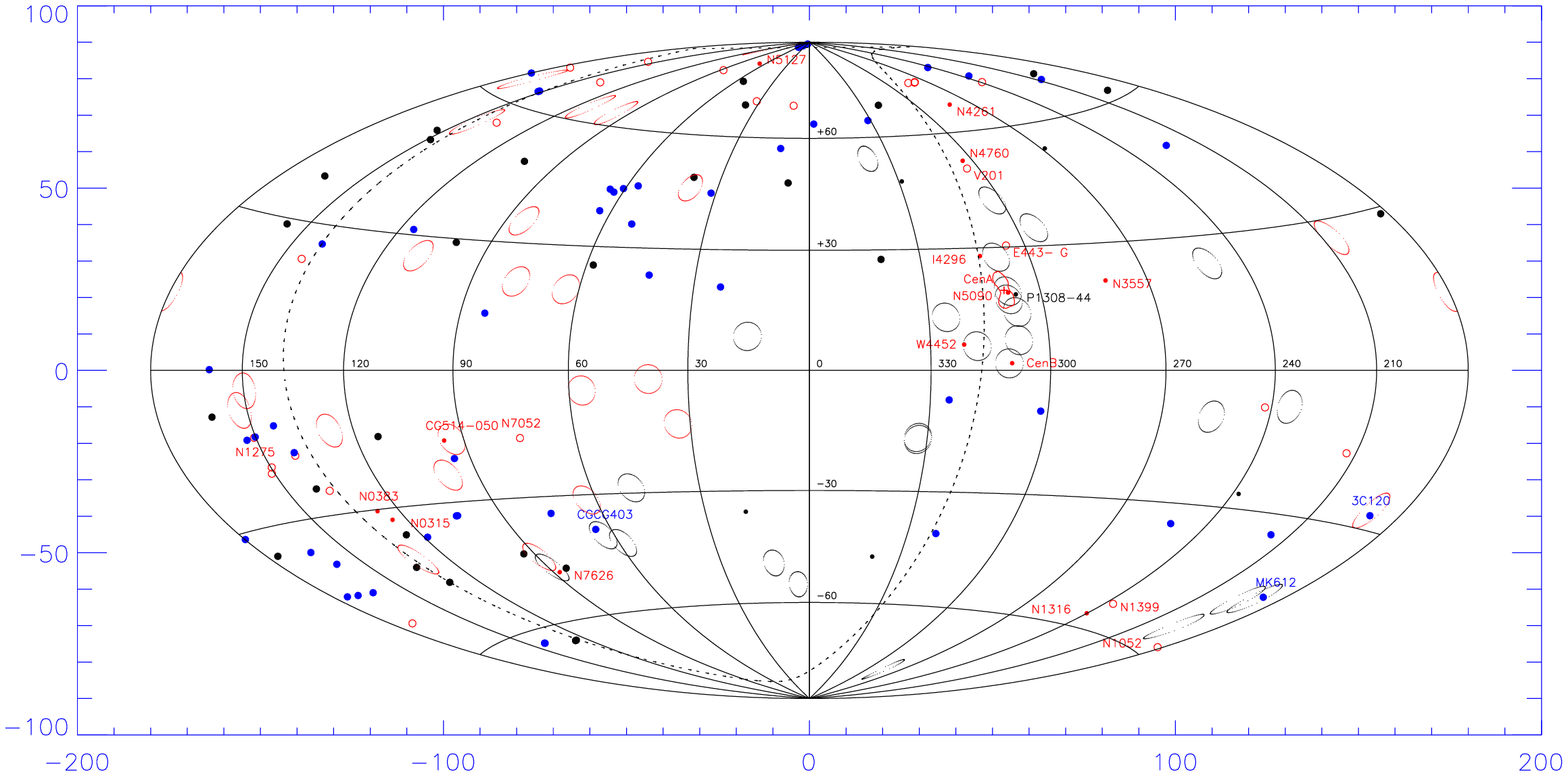}
\caption{A detail of Fig.~\ref{figjetned} showing the region around
the radiogalaxy Cen~A. Symbols are the same as in Fig.~\ref{figjetned}
except in the case of Cen~A, for which we use a cross to mark the galaxy 
center and a polygon to outline its radio jet structure as seen in a 403~MHz map.
}
\label{figsjetned}
\end{figure}

\begin{figure}
\includegraphics[width=4.5in,clip]{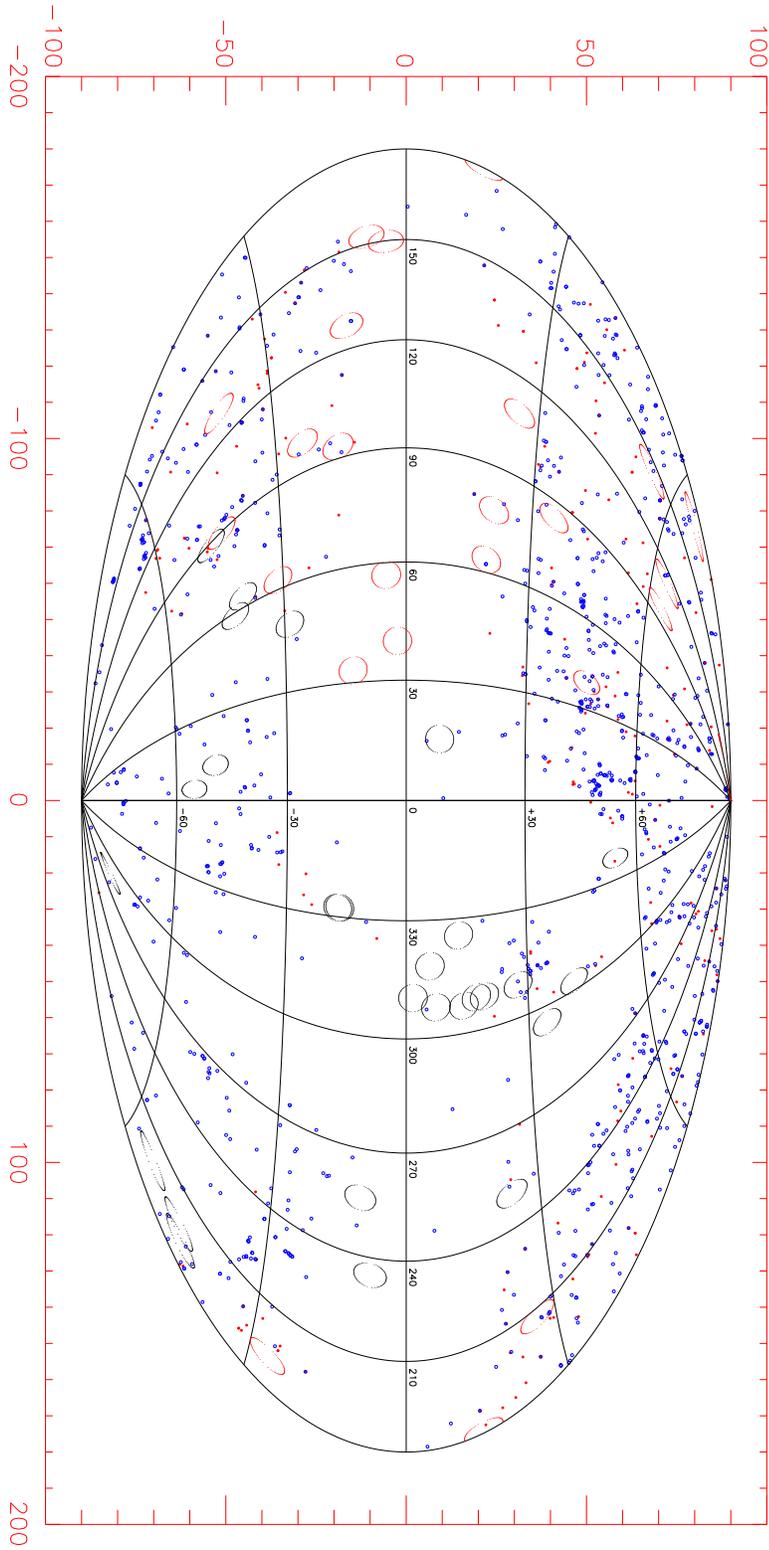}
\caption{A comparison of the arrival directions of the 27 UHECRs with
the positions of clusters of galaxies (as listed and defined by NED). 
The symbols for UHECRs are the same as in Fig.~\ref{figjetned}. 
Red symbols are used for galaxy clusters at 
D $\leq$ 75~Mpc, 
and blue symbols for galaxy clusters at
75~Mpc $<$ D $\leq$ 200~Mpc.
}
\label{figcluster}
\end{figure}

The results of the previous section were based only on 
UHECRs detected by the PAO. We have also tested for correlations
against the UHECRs with energy above 56~EeV detected by AGASA 
(red circles with radius 3.5\arcdeg\ in Fig.\ref{figjetned}. Among the 
nearby (D $\leq$ 75~Mpc) galaxies with extended radio jets, the only clear match
is to CGCG~514-050, a radiogalaxy at D=72.2~Mpc with radio extent
$\sim$250~kpc. The remaining three nearby galaxies with extended radio galaxies,
NGC~315, NGC~383, and NGC~5127, are not matched to within 3.5\arcdeg\ 
of any AGASA UHECR event, though there are several UHECRs at slightly
larger distances.  Additionally, One AGASA UHECR is close to the location of 
NGC~7626 and another is centered on the radiogalaxy 3C~120, at D=135~Mpc
and total radio extent $\sim$315~kpc.

Among the UHECRs detected by PAO, 
an isotropic distribution of the $\sim$20 UHECRs which are not in the
line of sight to nearby extended radiogalaxies cannot be ruled out at
high significance.
The significance of the correlation found by PA07 and PA08 between
UHECR arrival directions and nearby AGNs from the catalog
of V{\'e}ron-Cetty \& V{\'e}ron is much lower when one only
considers the $\sim$ 20 UHECRs not matched to nearby extended 
radiogalaxies and even lower when H~II nuclei in the
the AGN Catalog are deleted. The findings of PA07,PA08
were probably largely influenced by 
the correlation between UHECRs and nearby radiogalaxies
which we have shown above. 

The UHECR arrival directions are not strongly correlated
with either supernovae, extragalactic radio-supernovae,
or nearby groups and clusters of galaxies. 

The main difference between the results of the Pierre Auger
Collaboration and those from previous studies of UHECRs, e.g.
AGASA, could primarily be that the southern location of the
PAO is more privileged with respect
to nearby extended radiogalaxies.

\begin{acknowledgements}
NN dedicates this work to his thesis advisor and collaborator
Andrew S. Wilson. 
We acknowledge funding from ALMA 3105000 and 3016013, and
the FONDAP Center for Astrophysics.
This research has made use of the NASA/IPAC Extragalactic Database 
(NED) which is operated by the Jet Propulsion Laboratory, California 
Institute of Technology, under contract with the National Aeronautics 
and Space Administration.
\end{acknowledgements}

\clearpage
\onecolumn

\clearpage

\begin{figure}
\includegraphics[bb=0 20 595 842,width=\textwidth,clip]{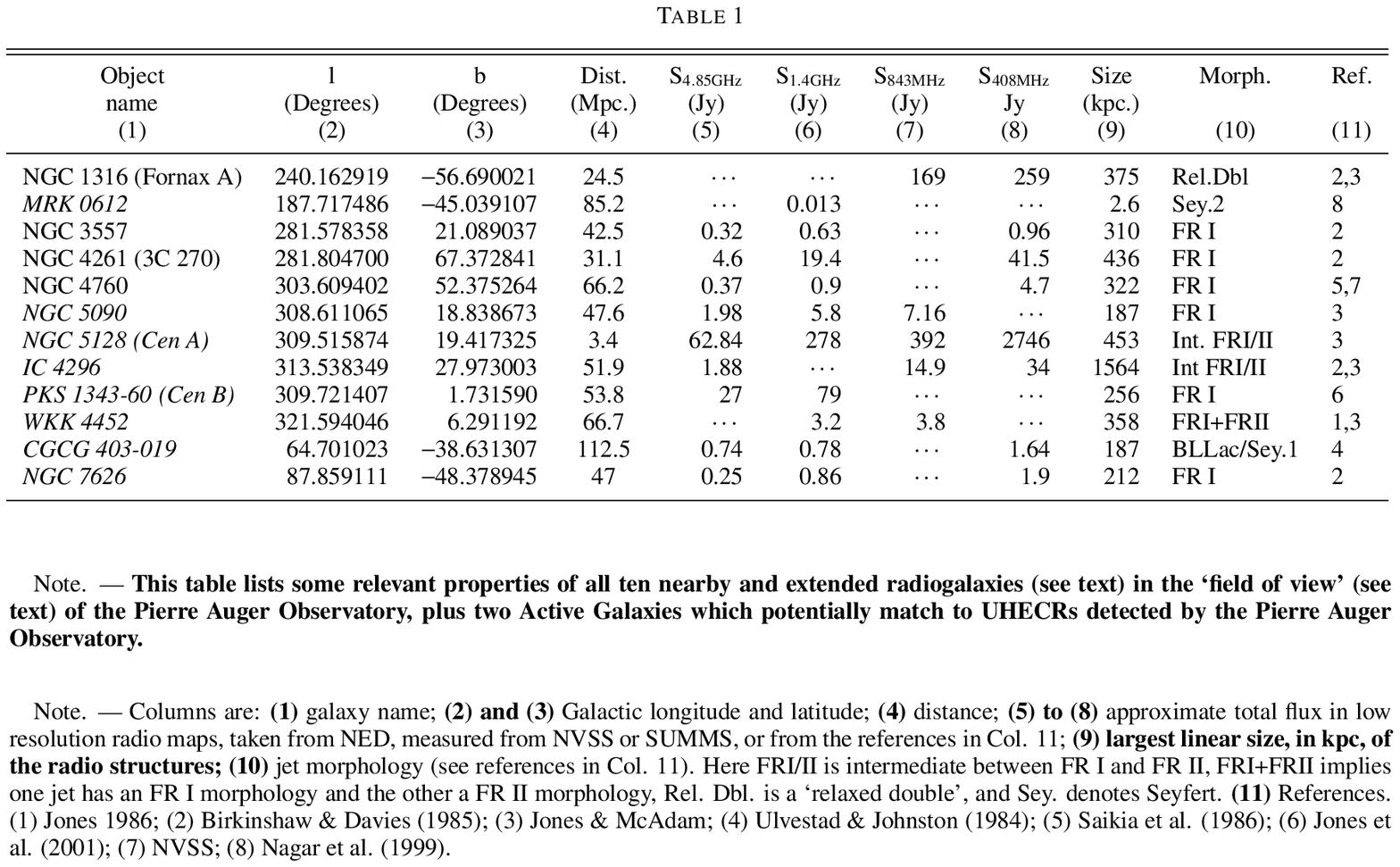}
\label{tab1}
\end{figure}

\end{document}